\documentclass{ws-ijmpa}
\usepackage[super,compress]{cite}
\usepackage{graphicx}
\usepackage{xcolor}

\begin{document}
\markboth{Ngo Phuc Duc Loc}{Sphaleron bound in some nonstandard cosmology scenarios}

%
\catchline{}{}{}{}{}
%

\title{Sphaleron bound in some nonstandard cosmology scenarios
}

\author{Ngo Phuc Duc Loc
}

\address{Department of Physics and Astronomy, University of New Mexico, Albuquerque, New Mexico 87131, USA\\
Department of Theoretical Physics, Vietnam National University, Ho Chi Minh City 700000, Vietnam}

\maketitle

\begin{history}
\received{Day Month Year}
\revised{Day Month Year}
\end{history}

\begin{abstract}
In the first scenario, we revise the upper bound of the cutoff scale of the dimension-six potential Higgs operator required for a successful electroweak baryogenesis in the case of a modified expansion history caused by the existence of a non-interacting scalar field at the time phase transition happens. The upper bound 860 GeV of the cutoff scale in the conventional case can be improved to 1 TeV in the modified expansion case under certain conditions. In the second scenario, we consider the Randall-Sundrum type II model. We show that the lower bound of the five-dimensional Planck scale in this model, which is determined from the validity of Newtonian gravitational potential at small distance, turns out to be crucial in eliminating this model as a viable candidate to satisfy the sphaleron bound; however, again modifying the expansion history by including a non-interacting scalar field at the electroweak scale can then make this model satisfy the sphaleron bound with a certain parameter space.

\end{abstract}

\ccode{PACS numbers: 11.15.Ex, 12.60.Fr, 98.80.Cq}


\section{Introduction}
If we assume that the Universe used to have an equal amount of matter and antimatter when it was born, then it is apparent that we need a kind of dynamical mechanism to break this symmetry to produce the observed Universe with more matter. Sakharov in 1967 has pointed out three conditions for such a mechanism \cite{sakharov}: violation of baryon number, violation of C and CP, and out-of-thermal equilibrium condition. Here, we focus on the third condition. It is known that this condition is not satisfied within the standard model of particle physics and cosmology and some modifications are needed.

In terms of the Standard Model of particle physics (SM), in an effort to minimally extend the model one usually includes some phenomenological terms in the Lagrangian using dimensional analysis. Among plenty of options, only the dimension-six operator for the Higgs potential, denoted here as $\mathcal{O}_6$, seems to be the most well-motivated one. The reason is that the form of the Higgs potential is rather weakly constrained by experiments and cannot be derived from pure theoretical considerations. The only thing we know is that there should be a minimum of the Higgs potential at zero temperature at the vacuum expectation value (VEV) around 246 GeV. The simplest possible option to obtain this form is to postulate the Higgs potential to be 
\begin{equation}
V(\phi)=\frac{\lambda}{4}(\phi^2-v^2)^2,
\end{equation}
where $\lambda>0$ is the quartic Higgs self-coupling and $v$ is the VEV. This leaves open questions whether or not this is the \textit{true} form of the Higgs potential. Therefore, it is possible to include the $\mathcal{O}_6$ operator of the form \cite{Ham}
\begin{equation}
\mathcal{O}_6=\frac{1}{8\Lambda^2}(\phi^2-v^2)^3,
\end{equation}
which preserves the minimum at VEV, where $\Lambda$ is the cutoff scale. This operator was very well-chosen so that it does not alter the three normalization conditions of the Higgs potential. This operator also directly changes the behavior of electroweak phase transition happening in the early Universe from second-order to first-order\footnote{It is also interesting to study the connection between sphaleron and  primordial gravitational waves generated from a first-order phase transition caused by the $\mathcal{O}_6$ operator as discussed in Ref. \citen{bian}.
}.

In terms of cosmology, we have little information about the transition between the inflationary era and the radiation era (for a good review, see Ref. \citen{allah}). It is possible that the usual standard radiation era \textit{did not} follow immediately after inflation. Therefore, we have a very fruitful direction of research for topics such as dark matter and reheating  in this transitional era. Our particular interest in this paper is to ask if such a \textit{nonstandard} transitional era can significantly affect the sphaleron bound of electroweak baryogenesis. As we will see, it does.

The final kind of possible modification we want to discuss is the possible existence of an infinite, curved extra spatial dimension and yet its physical deviation is not normally detectable. This is the Randall-Sundrum type II model (or RS2 for short) \cite{lisa}. This model predicts a deviation from the Newtonian gravitational potential at small distance as \cite{EPJC,lisa,roy}
\begin{equation}
V(r)=-\frac{Gm_1m_2}{r}\left(1+\frac{1024\pi^2M_4^4}{r^2M_5^6}\right),
\end{equation}
where $M_4$ and $M_5$ are four-dimensional and five-dimensional Planck scales, respectively. 
In Ref. \citen{EPJC}, we showed that the lower bound on $M_5$ is found by the tested validity of Newtonian gravitational law at small distance as
\begin{equation}\label{Newtonian lower bound of M_5}
M_5\gtrsim 3.84\times 10^9 \text{ GeV}.
\end{equation}

Given the above three different possible modifications of standard physics, we will organize this paper as follows. In a recent paper \cite{PRD}, we showed that there is an  upper bound on the cutoff scale of $\mathcal{O}_6$ in order to satisfy the out-of-thermal equilibrium condition of matter-antimatter asymmetry. In this paper, after reviewing the methodology of calculating sphaleron energy in Section \ref{2}, we will revise this upper bound in the scenario when the Friedmann equation is modified by including a non-interacting scalar field\footnote{When we say a non-interacting scalar field, we mean this field does not have non-gravitational interactions with ordinary matter.} in Section \ref{3}. Then, we will turn on to the Randall-Sundrum type II scenario in Section \ref{4} . The lower bound of $M_5$ mentioned above turns out to make this model invalid for electeoweak baryogenesis. Nevertheless, a modified expansion history again caused by a non-interacting scalar field in the electroweak era can make this model viable. This scenario is discussed in Section \ref{4}. We conclude in Section \ref{conclusions}. Natural units $\hbar=c=k_B=1$ are used throughout the paper.

\section{Sphaleron energy}\label{2}

The sphaleron rate is given by \cite{Gan}
\begin{equation}
\Gamma_{sph}\approx 5.68979\times 10^3\kappa \frac{v^7(T)}{T^6}exp\left(\frac{-E_{sph}(T)}{T}\right),
\end{equation}
where $10^{-4}\lesssim\kappa\lesssim 10^{-1}$ is the fluctuation determinant, $v(T)$ is the vacuum expectation value and $E(T)$ is the sphaleron energy corresponding to temperature $T$. The range of $\kappa$ we chose above is originally from Ref. \citen{Dine} and is also consistent with a lattice simulation conducted in Ref. \citen{lattice}, which gives the uncertainty of the sphaleron rate to be of order $\sim 10^3$. This sphaleron rate was calculated from the particle physics sector alone and is not affected by the expansion of the Universe. The sphaleron bound (also called decoupling condition\footnote{We will use these two terminologies interchangeably.}) required to preserve baryon asymmetry is
\begin{equation}\label{decoupling condition}
\Gamma_{sph}(T_c)<H(T_c),
\end{equation}
where $H$ is the Hubble rate at the time of electroweak phase transition and $T_c$ is the critical temperature.

Sphaleron energy at finite temperature is calculated as \cite{PRD}
\begin{equation}\label{E_sph general formula}
\tilde{E}_{sph}(T)=\frac{4\pi v}{g}\int_0^\infty d\xi \Bigg[4\left(\frac{df}{d\xi}\right)^2+\frac{8}{\xi^2}f^2(1-f)^2+\frac{\xi^2}{2}\left(\frac{dh}{d\xi}\right)^2+h^2(1-f)^2
+\frac{\xi^2}{g^2v^4}V_{eff}(h,T)\Bigg],
\end{equation}
where $V_{eff}(h,T)$ is the effective potential at temperature $T$, $f$ and $h$ are profile functions of sphaleron (also called radial functions) taking the following ansatz
\begin{equation}\label{ansatz f}
	f(\xi)=\begin{cases}
	\frac{\xi^2}{2a^2}, \hspace{1cm} \xi\leq a \,,\\
	1-\frac{a^2}{2\xi^2},\hspace{1cm} \xi\geq a ;
	\end{cases}
\end{equation}
\begin{equation}\label{ansatz h}
	h(\xi)=\begin{cases}
	\frac{4\xi}{5b},\hspace{1cm}\xi\leq b\,,\\
	1-\frac{b^4}{5\xi^4}\hspace{1cm}\xi\geq b\,.
	\end{cases}
\end{equation}
Note that $a$ and $b$ are scale-free parameters to be determined by minimizing the sphaleron energy functional in Eq. \ref{E_sph general formula}. The tilde symbol over the sphaleron energy means that $\xi$ and $h$ will need to undergo a rescaling prescription described below \textit{before} the sphaleron energy functional is minimized; the rescaled sphaleron energy is Eq. \ref{E(T)} below. The motivations for using this ansatz are discussed in Ref. \citen{PRD} and we will just very briefly repeat here: i) This ansatz satisfies the nonlinear equations of motion of the $SU(2)$ gauge theory in the asymptotic limit $\xi\rightarrow 0$ (sphaleron core); ii) This ansatz minimally ensures the convergence of sphaleron energy at finite temperature in the case of SM with the inclusion of $\mathcal{O}_6$; iii) This ansatz also generates results that are in good agreement with other numerical methods of calculations such as in Ref. \citen{Gan}.

\begin{figure}[h!]
\includegraphics[scale=1]{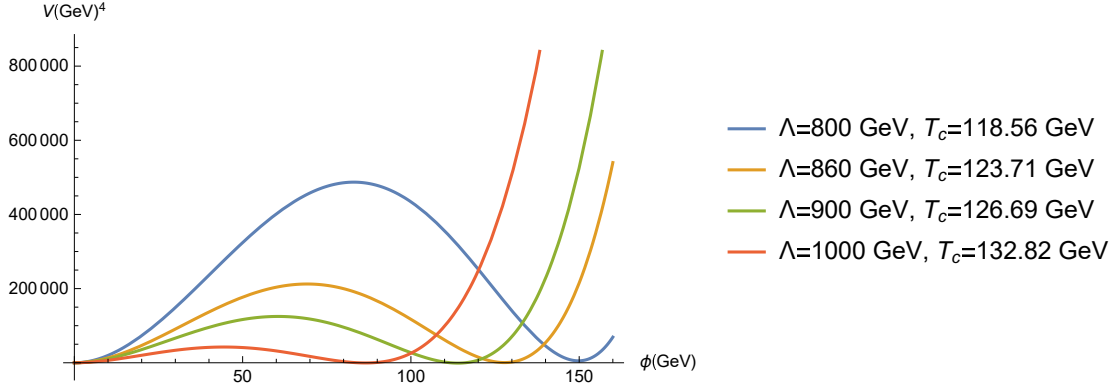}
\caption{The effective potential with various cutoff scales.}
\label{effective potential plot}
\end{figure}

In order to directly calculate sphaleron energy at finite temperature, we have to specify the effective potential. 
The effective potential at one-loop level of the Standard Model with the inclusion of $\mathcal{O}_6$ is \cite{PRD}
\begin{equation}\label{effective potential}
V_{eff}(\phi,T)=\frac{\lambda (T)}{4}\phi^4-ET\phi^3+D(T^2-T_0^2)\phi^2+\Lambda (T)+\frac{1}{8\Lambda ^2}(\phi^2-v^2)^3,
\end{equation}
where
\begin{equation}\label{lambda(T)}
\lambda(T)=\frac{m_h^2}{2v^2}-\frac{1}{16\pi^2v^4}\sum_{i=h,W,Z,t}n_im_i^4ln\left(\frac{m_i^2}{A_iT^2}\right),
\end{equation}
\begin{equation}\label{T_0^2}
T_0^2=\frac{1}{D}\left[\frac{m_h^2}{4}-\frac{1}{32\pi^2v^2}\sum_{i=h,W,Z,t}n_im_i^4\right],
\end{equation}
\begin{equation}\label{Lambda(T)}
\Lambda(T)=\frac{m_h^2v^2}{8}-\frac{1}{128\pi^2}\sum_{i=h,W,Z,t}n_im_i^4-\frac{41\pi^2T^4}{180},
\end{equation}
\begin{equation}\label{E}
E\equiv\frac{1}{12\pi v^3}\sum_{i=h,W,Z}n_im_i^3=\frac{m_h^3+6m_W^3+3m_Z^3}{12\pi v^3},
\end{equation}
\begin{equation}\label{D}
D\equiv\sum_{i=h,W,Z}\frac{n_im_i^2}{24v^2}-\frac{n_tm_t^2}{48v^2}=\frac{m_h^2+6m_W^2+3m_Z^2+6m_t^2}{24v^2},
\end{equation}
and $ln(A_B)=3.9076$ for bosons,  $ln(A_F)=1.1351$ for fermions. 
Note that the form of $\mathcal{O}_6$ we chose does not alter the three normalization conditions: $V(v)=0, V'(v)=0,V''(v)=m_h^2$. The effective potential of SM is recovered in the limit $\Lambda\rightarrow\infty$. The effective potential in Eq. \ref{effective potential} is plotted in Fig. \ref{effective potential plot} showing that we have a first-order electroweak phase transition.

Define $s\equiv v(T)/v$, we can rewrite the effective potential as follows:
\begin{equation}
\begin{aligned}
V_{eff}(h,T)&=\frac{\lambda (T)}{4}\phi^4-ET\phi^3+D(T^2-T_0^2)\phi^2+\Lambda (T)+\frac{1}{8\Lambda ^2}(\phi^2-v^2)^3\\
&= \frac{\lambda (T)v^4}{4}s^4h^4-ETv^3s^3h^3+D(T^2-T_0^2)v^2s^2h^2+\frac{v^6}{8\Lambda ^2}(s^2h^2-1)^3+\Lambda (T)\\
&=\frac{\lambda (T)v^4}{4}s^4(h^4-1)-ETv^3s^3(h^3-1)+D(T^2-T_0^2)v^2s^2(h^2-1)\\
&+\frac{v^6}{8\Lambda ^2}\Big[(s^6(h^6-1)-3s^4(h^4-1)+3s^2(h^2-1)\Big]+C(T).
\end{aligned}
\end{equation}
In the first line, we only rewrite the effective potential in Eq. \ref{effective potential}. In the second line, we substitute $\phi=vh$ and rescale the radial function $h$ as $h\rightarrow sh$. In the third line, we add and remove some temperature-dependent constants and set the rest to be $C(T)$. The sphaleron energy will be rewritten as
\begin{equation}\label{E(T)}
\begin{aligned}
E_{sph}(T)&=\frac{4\pi v}{gs}\int_0^\infty d\xi \Bigg\{s^2\Bigg[4\left(\frac{df}{d\xi}\right)^2+\frac{8}{\xi^2}f^2(1-f)^2+\frac{\xi^2}{2}\left(\frac{dh}{d\xi}\right)^2+h^2(1-f)^2\Bigg]\\
&+\frac{\xi^2}{g^2s^2}\Bigg[\frac{\lambda (T)}{4}s^4(h^4-1)-\frac{ET}{v} s^3(h^3-1)+\frac{D(T^2-T_0^2)}{v^2} s^2(h^2-1)\\
&+\frac{v^2}{8\Lambda ^2}\Big[(s^6(h^6-1)-3s^4(h^4-1)+3s^2(h^2-1)\Big]\Bigg]\Bigg\},
\end{aligned}
\end{equation}
where we have ignored the irrelevant $C(T)$ and also rescaled $\xi\rightarrow \xi/s$. This whole procedure was given in our previous work in Ref. \citen{PRD}.

Now, we just need to insert the profile functions ansatz in Eqs. \ref{ansatz f} and \ref{ansatz h} into the sphaleron energy functional in Eq. \ref{E(T)}, minimize this formula with respect to the two parameters $a$ and $b$, and  we will get the  sphaleron energy at the critical temperature $T_c$ as shown in Table \ref{table of sphaleron energy}. We are interested in the critical temperature because at this temperature the potential energy of two minima are equal and the vacuum starts to occupy the new minimum. If the sphaleron bound in Eq. \ref{decoupling condition} is satisfied at this temperature, then it will always be satisfied later on because as the temperature decreases, the sphaleron energy will increase and therefore the sphaleron rate will decrease exponentially, while the Hubble rate will only decrease with a polynomial temperature dependence. Some authors might want to study this condition at the bubble nucleation temperature instead, but generally the bubbles of the new true vacuum can only nucleate when the phase transition actually occurs, and that can only happen at or below the critical temperature. 

\begin{table}[h]
\tbl{Sphaleron energy calculated in units of $4\pi v/g\approx 4.738$ TeV with different values of the cutoff scale. The SM case is recovered in the limit $\Lambda\rightarrow\infty$.}{
	\begin{tabular}{|c|c|c|c|c|c|}
		\hline\hline
	$\Lambda(GeV)$ & $T_c(GeV)$	&  $v_c(GeV)$ & $E(T_c)$ & a & b
		\\
		\hline
		800 & 118.56 & 149.78 & 1.136 & 2.738 & 3.272\\
		\hline
		860 & 123.71 & 127.81 & 0.966 & 2.762 & 3.342\\
		\hline
		900 & 126.69 & 113.98 & 0.860 & 2.768 & 3.359\\
		\hline
		1000 & 132.82 & 86.53 & 0.653 & 2.769 & 3.364\\
		\hline
		$\infty$& 158.376 & 35.346 & 0.271 & 2.673 & 3.086\\ 
		\hline\hline
	\end{tabular}
	\label{table of sphaleron energy}}
\end{table}

\section{Sphaleron bound in 4D spacetime with modified expansion rate}\label{3}

The radiation era is usually thought to follow immediately after the inflationary era with the Friedmann equation as
\begin{equation}\label{H_rad}
H^2_r=\frac{8 \pi}{3 M_{4}^2}\rho_r,
\end{equation}
and the energy density of radiation is
\begin{equation}\label{rho_r}
\rho_r=\frac{\pi^2}{30}g_*T^4,
\end{equation}
where $M_{4}=1.22\times 10^{19}$ GeV is the usual four-dimensional Planck scale and $g_*=106.75$ is the number of effective relativistic degrees of freedom corresponding to temperature $T$. We will also find the energy density of radiation at BBN useful
\begin{equation}\label{rho_r BBN}
\rho_r^{BBN}=\frac{\pi^2}{30}g_{*}^{BBN} T^4_{BBN},
\end{equation}
where $g_*^{BBN}=10.75$ is the number of effective relativistic  degrees of freedom at BBN and $T_{BBN}\approx 1$ MeV.

However, this is not necessarily true since we do not have much observational information after the inflationary era but prior to Big Bang Nucleosynthesis (BBN). Thus there is some ambiguity in this era and the Hubble rate in this era may take a different form from the standard one. We would like to use BBN as the observational constraint on this possible deviation as \cite{JCAP}
\begin{equation}
\frac{H}{H_r}\Bigg|_{BBN}=\sqrt{1+\frac{7}{43}\Delta N_{\nu eff}},
\end{equation}
where $H$ is a new modified Hubble rate and $\Delta N_{\nu eff}$ is the difference between the calculated and observed effective number of neutrino species. The notation on the left hand side means that the quantity is calculated at BBN. With the observed value of $\Delta N_{\nu eff}<0.16$ \cite{BBN}, we have the constraint 
\begin{equation}\label{BBN constrain}
\frac{H}{H_r}\Bigg|_{BBN}<1.0129.
\end{equation}

The assumption we make is that electroweak phase transition happens after inflation but before BBN in a nonstandard expansion era. The purpose is to make the model satisfy the decoupling condition with a nonstandard Hubble rate at the electroweak scale. Assume that such a modification is caused by the coexistence of a scalar field such as dark matter \cite{lew} or inflaton\footnote{If the scalar field is inflaton, a nonstandard reheating mechanism is needed because the kination era does not have an oscillating behavior that we expect for a conventional reheating.} \cite{joyce}, which do not have non-gravitational interactions with normal matter and usual radiation, then the modified Friedmann equation is
\begin{equation}\label{H_ew}
H^2_{ew}= \frac{8\pi}{3M_4^2}(\rho_\phi+\rho_r).
\end{equation}
The notation $H_{ew}$ stands for the modified Hubble rate at the electroweak scale. From Eq. \ref{H_ew} and Eq. \ref{H_rad}, we have the ratio
\begin{equation}\label{ratio of H_ew and H_r}
\left(\frac{H_{ew}}{H_r}\right)^2=\frac{\rho_\phi}{\rho_r}+1.
\end{equation}
We also have the following scaling laws of energy densities
\begin{equation}\label{scaling law of radiation}
\rho_r=\rho_r^{BBN}\left(\frac{a_{BBN}}{a}\right)^4,
\end{equation}
\begin{equation}\label{scaling law of scalar field}
\rho_\phi=\rho_\phi^{BBN}\left(\frac{a_{BBN}}{a}\right)^n.
\end{equation}
Although the case with $\rho_\phi\sim\frac{1}{a^6}$ seems to be the most well-motivated case \cite{lew,joyce}, we will keep the calculations general. Also note that we \textit{cannot} use the temperature scaling law $T\sim 1/a$ from the electroweak era way down to the BBN era because it is invalid when particles already started to decouple from the thermal bath well before BBN. Eq. \ref{ratio of H_ew and H_r} then becomes
\begin{equation}
\begin{aligned}
\left(\frac{H_{ew}}{H_r}\right)^2&= \frac{\rho_\phi^{BBN}}{\rho_r^{BBN}}\left(\frac{a_{BBN}}{a}\right)^{n-4}+1\\
&= \left[\left(\frac{H_{ew}}{H_r}\Bigg|_{BBN}\right)^2-1\right]\left(\frac{\rho_r}{\rho_r^{BBN}}\right)^{\frac{n-4}{4}}+1\\
&= \left[\left(\frac{H_{ew}}{H_r}\Bigg|_{BBN}\right)^2-1\right] \left(\frac{g_*}{g_*^{BBN}}\right)^{\frac{n-4}{4}}\left(\frac{T}{T_{BBN}}\right)^{n-4}+1,
\end{aligned}
\end{equation}
where in the first line we used Eq. \ref{scaling law of radiation} and Eq. \ref{scaling law of scalar field}, in the second line we used Eq. \ref{scaling law of radiation} and Eq. \ref{ratio of H_ew and H_r} calculated at BBN, in the third line we used Eq. \ref{rho_r} and Eq. \ref{rho_r BBN}. Finally, using Eq. \ref{H_rad} and Eq. \ref{rho_r} we have the modified Hubble rate at the electroweak scale as
\begin{equation}\label{H_ew final}
H_{ew}=\Bigg\{\left[\left(\frac{H_{ew}}{H_r}\Bigg|_{BBN}\right)^2-1\right] \left(\frac{g_*}{g_*^{BBN}}\right)^{\frac{n-4}{4}}\left(\frac{T}{T_{BBN}}\right)^{n-4}+1\Bigg\}^{1/2}\Bigg\{\frac{4 \pi^3}{45 M_4^2}g_*T^4\Bigg\}^{1/2}.
\end{equation}
This equation is very useful since we can directly impose the observational constraint in Eq. \ref{BBN constrain} on the unknown $H_{ew}$. 

\begin{figure}[h!]
\centering
\includegraphics[scale=1]{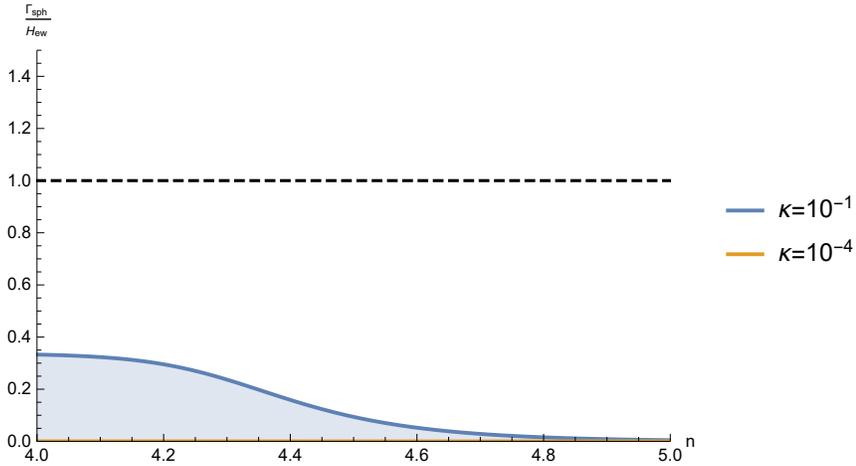}
\caption{$\Lambda=800$ GeV. Decoupling condition is satisfied for all $n$ when $\Lambda=800$ GeV.}
\label{cutoff800}
\end{figure}

\begin{figure}[h!]
\centering
\includegraphics[scale=1]{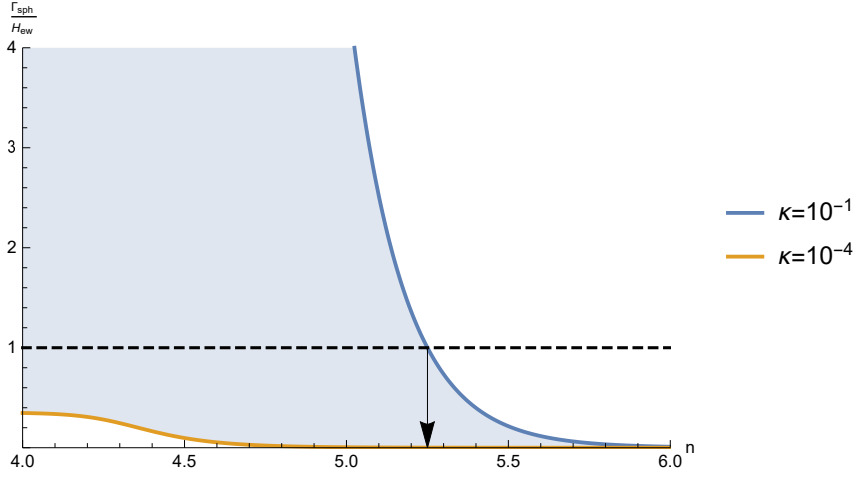}
\caption{$\Lambda=860$ GeV. If the sphaleron rate is maximal, decoupling condition can only be satisfied with $n\gtrsim 5.25$ when $\Lambda=860$ GeV.}
\label{cutoff860}
\end{figure}

\begin{figure}[h!]
\centering
\includegraphics[scale=1]{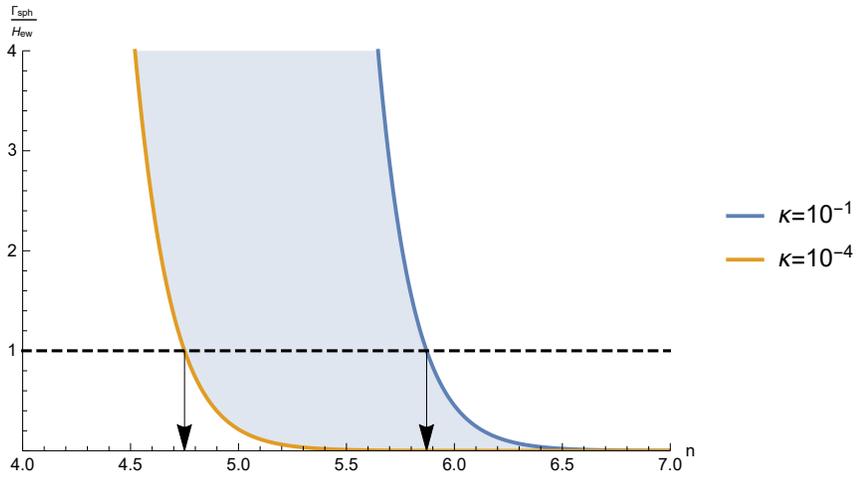}
\caption{$\Lambda=900$ GeV. Depending on the sphaleron rate, decoupling condition is satisfied either with  $n\gtrsim 4.75$ or $n\gtrsim 5.87$ when $\Lambda=900$ GeV.} 
\label{cutoff900}
\end{figure}

\begin{figure}[h!]
\centering
\includegraphics[scale=1]{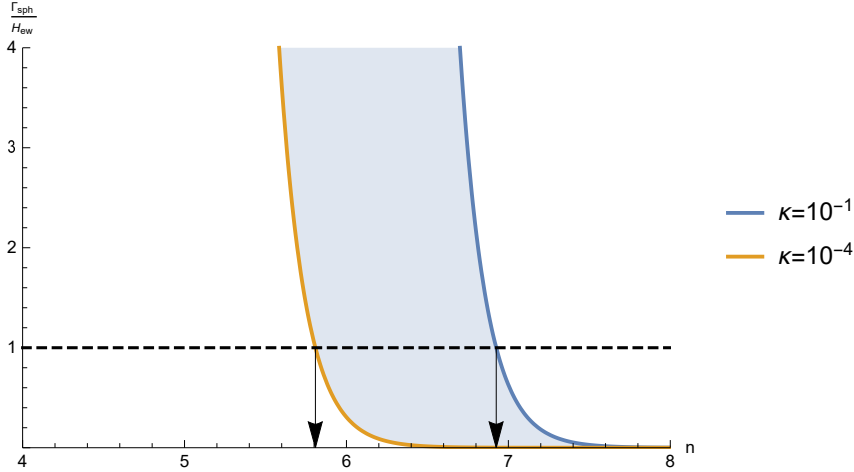}
\caption{$\Lambda=1000$ GeV. Depending on the sphaleron rate, decoupling condition is satisfied either with $n\gtrsim 5.81$ or $n\gtrsim 6.93$ when $\Lambda=1000$ GeV.}
\label{cutoff1000}
\end{figure}

\begin{figure}[h!]
\centering
\includegraphics[scale=1]{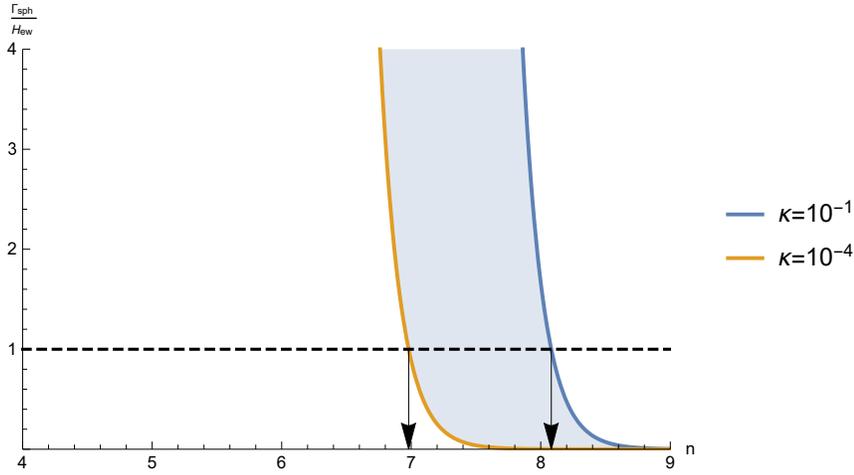}
\caption{Standard model case. Depending on the sphaleron rate, decoupling condition is satisfied either with $n\gtrsim 6.98$ or $n\gtrsim 8.08$ when $\Lambda=\infty$ (SM case).}
\label{SM}
\end{figure}

From the calculated sphaleron energy in the previous section, let's consider the requirement for the decoupling condition $\Gamma_{sph}(T_c)<H_{ew}(T_c)$ to be satisfied with different values of the cutoff scale, as shown in Figures \ref{cutoff800}, \ref{cutoff860}, \ref{cutoff900}, \ref{cutoff1000} and \ref{SM}.

We have the following remarks:
\begin{itemize}
\item When $\Lambda=800$ GeV, the contribution of $\mathcal{O}_6$ is significant and hence the decoupling condition is satisfied irrespective of $n$.
\item $\Lambda=860$ GeV is roughly the upper bound found in \cite{PRD} with the standard radiation era when we chose the minimal sphaleron rate. In the modified expansion scenario, we would need $n\gtrsim 5.25$ if the sphaleron rate is  maximal (with $\kappa=10^{-1}$).
\item When $\Lambda=900$ GeV, the standard $n=4$ case is not appropriate anymore and the sphaleron bound is guaranteed only if $n\gtrsim 4.75$ or $n\gtrsim 5.87$, depending on the sphaleron rate. The well-motivated option with $n=6$ can make sure that the sphaleron bound is satisfied irrespective of $\kappa$.
\item The case with $\Lambda=1$ TeV can still satisfy sphaleron bound with the well-motivated case $n=6$ if the sphaleron rate is sufficiently small (with $\kappa$ is somewhere near $10^{-4}$).
\item The SM case (when $\Lambda\rightarrow\infty$) requires at least $n= 7$.
\end{itemize}

The qualitative observation is that, as the power $n$ increases, the Hubble rate becomes larger and larger until the decoupling condition is satisfied. This fact can easily be seen mathematically from Eq. \ref{H_ew final} and can also be expected from the physical intuition that the presence of an extra scalar field at the electroweak scale would contribute its energy density to the Hubble rate until it becomes larger than the sphaleron rate, in such a way that its energy density decreases quickly enough to match the observational constraint in the BBN era.

\section{Sphaleron bound in Randall-Sundrum type II model}\label{4}
There is a possibility to realize a modified Friedmann equation as a consequence of an extra infinite and curved spatial dimension, which is the Randall-Sundrum type II model (RS2).
The modified Friedmann equation of the RS2 model is \cite{roy}
\begin{equation}\label{H_RS}
H_{RS}^2=\frac{8\pi}{3M_4^2}\rho\left(1+\frac{\rho}{2\lambda}\right),
\end{equation}
where $\lambda$ is the brane tension relating the four-dimensional and five-dimensional Planck scales as \cite{roy}
\begin{equation}
\lambda=\frac{3M_5^6}{4\pi M_4^2}.
\end{equation}
The notation $H_{RS}$ stands for the modified Hubble rate in RS2 model at the electroweak scale. We will show in subsection \ref{standard RS2} that this model fails to satisfy the sphaleron bound. We will then show in subsection \ref{modified RS2}, however, that an effort to extend the model by again introducing a new non-interacting scalar field can help. From now on, we will only consider sphaleron rate calculated in the SM case, partly because the decoupling condition can be already satisfied in the $SM+\mathcal{O}_6$  case and no further bounds on $M_5$ can be imposed, and partly because finding a parameter space that can satisfy the decoupling condition in the SM case itself is particularly compelling.
\subsection{Standard RS2 model}\label{standard RS2}
If the energy density of the Universe at the time of electroweak phase transition only contains radiation, then from Eq. \ref{H_rad} and Eq. \ref{H_RS} we have
\begin{equation}
\left(\frac{H_{RS}}{H_r}\right)^2\Bigg|_{BBN}=\frac{\rho_r^{BBN}}{2\lambda}+1.
\end{equation}
From the observational constraint in Eq. \ref{BBN constrain}, we have 
\begin{equation}
M_5> 5.91\times 10^4 \text{ GeV}.
\end{equation}
This lower bound of $M_5$ is complied with the lower bound discussed in Eq. \ref{Newtonian lower bound of M_5}.

From the calculated sphaleron energy of the SM model in Table \ref{table of sphaleron energy},   we can find the upper bound of $M_5$ from the decoupling condition as
\begin{equation}
\Gamma_{sph}<H_{RS}=\sqrt{\left(\frac{\rho_r}{2\lambda}+1\right)\left(\frac{8\pi}{3M_4^2}\rho_r\right)},
\end{equation}
with the largest possible upper bound corresponding to $\kappa=10^{-4}$ is
\begin{equation}
M_5<4.98\times 10^5 \text{ GeV}.
\end{equation}
This upper bound violates the lower bound mentioned in Eq. \ref{Newtonian lower bound of M_5}, so this model cannot work. 

\subsection{RS2 model with modified expansion rate}\label{modified RS2}
If we modify the expansion rate of RS2 model in the electroweak era by including a new non-interacting scalar field, then from Eq. \ref{H_RS} we have
\begin{equation}
H_{RS}^2=\frac{8\pi}{3M_4^2}(\rho_\phi+\rho_r)+\frac{4\pi}{3M_4^2\lambda}(\rho_\phi+\rho_r)^2.
\end{equation}
Taking the ratio of the above equation with the one of standard cosmology in Eq. \ref{H_rad} we get
\begin{equation}\label{HRS/Hr}
\left(\frac{H_{RS}}{H_r}\right)^2= \frac{\rho_\phi}{\rho_r}+1+\frac{\rho_r}{2\lambda}\left(\frac{\rho_\phi}{\rho_r}+1\right)^2.
\end{equation}
The only possible non-negative solution to the above equation is 
\begin{equation}\label{delta}
\delta\equiv\frac{\rho_\phi}{\rho_r}=\left[\sqrt{1+\frac{2\rho_r}{\lambda}\left(\frac{H_{RS}}{H_r}\right)^2}-1\right]\frac{\lambda}{\rho_r}-1,
\end{equation}
where we used the notation $\delta$ to denote the ratio between the energy density of the scalar field and the usual radiation for future convenience. The positivity requirement is
\begin{equation}
\left(\frac{H_{RS}}{H_r}\right)^2>\frac{\rho_r}{2\lambda}+1,
\end{equation}
which not surprisingly implies an identical lower bound on $M_5$ as in the previous subsection.

Following the same procedure of Section \ref{3}, Eq. \ref{HRS/Hr} becomes
\begin{equation}
\begin{aligned}
\left(\frac{H_{RS}}{H_r}\right)^2&= \frac{\rho_\phi^{BBN}}{\rho_r^{BBN}}\left(\frac{a_{BBN}}{a}\right)^{n-4}+1+\frac{\rho_r}{2\lambda}\left[\frac{\rho_\phi^{BBN}}{\rho_r^{BBN}}\left(\frac{a_{BBN}}{a}\right)^{n-4}+1\right]^2\\
&= \delta_{BBN}\left(\frac{\rho_r}{\rho_r^{BBN}}\right)^{\frac{n-4}{4}}+1+\frac{\rho_r}{2\lambda}\left[\delta_{BBN}\left(\frac{\rho_r}{\rho_r^{BBN}}\right)^{\frac{n-4}{4}}+1\right]^2,
\end{aligned}
\end{equation}
where in the first line we used Eq. \ref{scaling law of radiation} and Eq. \ref{scaling law of scalar field}, in the second line we used Eq. \ref{scaling law of radiation} and Eq. \ref{delta} calculated at BBN. So finally we have
\begin{equation}
H_{RS}=\Bigg\{\delta_{BBN}\left(\frac{\rho_r}{\rho_r^{BBN}}\right)^{\frac{n-4}{4}}+1+\frac{\rho_r}{2\lambda}\left[\delta_{BBN}\left(\frac{\rho_r}{\rho_r^{BBN}}\right)^{\frac{n-4}{4}}+1\right]^2\Bigg\}^{1/2}\Bigg\{\frac{8\pi}{3M_4^2}\rho_r\Bigg\}^{1/2}.
\end{equation}
This equation plays a similar role to Eq. \ref{H_ew final} in the previous section but now we have an additional parameter $M_5$ characterizing the RS2 model, which is implicitly contained in $\lambda$. Note that the temperature dependence of $\rho_r$ in Eq. \ref{rho_r} and the BBN observational constraint (Eq. \ref{BBN constrain}) on $\delta_{BBN}$ are understood.

As mentioned in the introduction, we only have the lower bound of $M_5$ obtained from the tested validity of Newtonian gravitational potential at small distance, while the upper bound has never been determined by any methods before. Therefore, we want to make an estimation of the upper bound of $M_5$ by requiring that this model complies with the sphaleron bound. In order to be more specific and conclusive, it is necessary to choose the power $n$ of $\rho_\phi\sim a^{-n}$ in a well-motivated case such as $n=6$. In the SM case (when $\Lambda\rightarrow\infty$), the decoupling condition $\Gamma_{sph}(T_c)<H_{RS}(T_c)$ implies the largest possible upper bound of $M_5$ to be 
\begin{equation}
M_5\lesssim 1.03\times 10^{14} \text{ GeV},
\end{equation}
which corresponds to $\kappa=10^{-4}$. It is straightforward to check that smaller values of $n$ (such as $n=5$) would require smaller upper bound of $M_5$, which in turn would violate the lower bound of $M_5$; larger values of $n$ (such as $n=7$) would imply larger upper bounds of $M_5$ that will eventually exceed the 4D Planck scale, which is not very encouraging.

So the final range of $M_5$ is
\begin{equation}
3.84\times 10^9 \text{ GeV} \lesssim M_5\lesssim 1.03\times 10^{14} \text{ GeV}.
\end{equation}
This range can be further tightened either from the lower bound by future high-precision measurements of the Newtonian gravitational law at small distance or from the upper bound by reducing uncertainty in the calculations of the sphaleron rate. We want to highlight that the above parameter space guarantees that the decoupling condition is satisfied \textit{within} the context of SM itself. Unlike the 4D spacetime case, which would require a novel model to realize at least $n=7$, the RS2 model only requires $n=6$, which is already realized in many well-motivated theories.

Finally, we should also note that although we discussed the RS2 model specifically, the calculations were completely general and are applicable to any Friedmann equations with the  quadratic correction of energy density, with the cost of introducing a new parameter ($M_5$ in the particular case of RS model). 

\section{Conclusions}\label{conclusions}
In this paper, we showed that the presence of a new scalar field in the transitional era between the inflationary era and the usual radiation era can have significant impacts on the sphaleron condition of electroweak baryogenesis. Our current work is basically an extension of our earlier work in Ref. \citen{PRD}. The upper bound of the cutoff scale of the dimension-six potential Higgs operator, which is around $860$ GeV in the standard radiation era \cite{PRD}, can be improved to larger values if the power $n$ in the scaling law $\rho_\phi\sim 1/a^n$ of the new non-interacting scalar field takes larger values. In the most well-motivated case of $n=6$, the cutoff scale can be as large as $\Lambda=1$ TeV if the sphaleron rate is close to minimum with $\kappa=10^{-4}$. The raise of the upper bound of the cut-off scale is encouraged due to tight constraints of higher-dimensional operators in current experiments. Meanwhile, the Standard Model (when $\Lambda\rightarrow\infty$) would require at least $n= 7$, which we have not seen any models that can realize this. It is interesting to find out if there is such a theory.

In the case of standard Randall-Sundrum type II model, it is not possible to satisfy the sphaleron bound due to violation of the lower bound of the five-dimensional Planck scale $M_5$ obtained from the validity of Newtonian gravitational law at small distance. However, this model \textit{can} satisfy the sphaleron bound if we again add a new non-interacting scalar field to the model. The interesting feature is that the sphaleron bound can then be satisfied within the Standard Model of particle physics itself if $M_5$ is in the range $3.84\times 10^9 \text{ GeV} \lesssim M_5\lesssim 1.03\times 10^{14} \text{ GeV}$ when $n=6$ and $\kappa=10^{-4}$. The upper bound is the largest possible bound. Future accurate experimental tests of Newtonian gravitational potential at small distance can only raise the lower bound, while more accurate calculations of the sphaleron rate can only lower the upper bound. It is possible that we can tighten it to a single value. We can think of $M_5$ as a new fundamental energy scale of gravity in the more fundamental 4+1 dimensional spacetime (the bulk), and the usual Planck scale in 3+1 dimensional spacetime ($M_4$) is just an effective energy scale of gravity (on the brane). In this sense, demanding the model to satisfy the sphaleron bound under certain well-motivated assumptions has shed some insights on the study of fundamental physics.

We can add some more comments about the visions beyond this paper. The dimension-six Higgs operator we considered is a phenomenological term that was added to the effective  Lagrangian to parameterize the unknown UV complete theory. It did a very good job to make the model satisfy the sphaleron bound. But we are not aware of any specific UV models that can realize this operator in the low energy limit. However, we note that higher dimensional Higgs operators can generally be induced in some UV complete theories as discussed in Ref. \citen{UV}. Examples are the existence of extra scalars with strong couplings to the Higgs boson \cite{UV1,UV2} or the composite Higgs model in which the Higgs arises as a pseudo-Goldstone boson of a global SO(5) symmetry broken down to SO(4) \cite{UV3}. In terms of the modified expansion scenario, the review article Ref. \citen{allah} has a fairly comprehensive collection of various possibilities. A typical example of a nonstandard expansion phase is kination in which the kinetic energy term of a scalar field dominates its potential energy, and $\rho_\phi\sim a^{-6}$ is realized  \cite{kination}. Such a modification can in turn lead to some changes of inflationary predictions. As pointed out in Ref. \citen{allah}, the general consequence of a nonstandard expansion phase on predictions of inflation is encoded in the remaining number of e-folds $N_*$ at the time of horizon exit. It can be modified from the traditional range $N_*\simeq 50-60$ to the new range $N_*\simeq 30-60$. This can actually accommodate new (possibly more complicated) models of inflation but can also eliminate some traditional ones. Because we do not have observational information for the era prior to BBN, it is therefore important to be open-minded for such a complicated scenario.

\end{document}